\documentclass[aps,prl,twocolumn,superscriptaddress,letter]{revtex4}
\usepackage{amssymb}
\usepackage{color}
\usepackage{graphicx}
\usepackage{dcolumn}
\usepackage{bm}

\begin{document}

\title{Quantum Non-Magnetic state near Metal-Insulator Transition - a
Possible Candidate of Spin Liquid State}
\author{Gao-Yong Sun}
\affiliation{Department of Physics, Beijing Normal University, Beijing 100875, China}
\author{Su-Peng Kou}
\thanks{Corresponding author}
\email{spkou@bnu.edu.cn}
\affiliation{Department of Physics, Beijing Normal University, Beijing 100875, China}

\begin{abstract}
In this paper, based on the formulation of O(3) non-linear $\sigma $ model,
we study two dimensional $\pi -flux$ Hubbard model at half-filling. A
quantum non-magnetic insulator is explored near metal-insulator transition
that may be a possible candidate of spin liquid state. Such quantum
non-magnetic insulator on square lattice is not induced by frustrations.
Instead, it originates from quantum spin fluctuations with relatively small
effective spin-moments. In the strong coupling limit, our results of spin
velocity and spin order parameter agree with results obtained from earlier
calculations.

\end{abstract}

\maketitle

People have been seeking for quantum spin liquid states in spin models with
predominantly antiferromagnetic short ranged interactions for over two
decades \cite{Fazekas}. For example, various approaches show that quantum
spin liquids may exist in two-dimensional (2D) $S=1/2$ \textrm{J}$_{1}$%
\textrm{-J}$_{2}$ model or Heisenberg model on Kagom\'{e} lattice. In these
models, the quantum spin liquids are accessed (in principle) by appropriate
frustrating interactions. In particular, such type of spin liquid states can
be described by the Hubbard model formalism in the strong coupling limit.

Recent experiments on the triangular lattice show that the spin liquid
ground state may be realized in the organic material $\mathrm{\kappa
-(BEDT-TTF)}_{2}\mathrm{CU}_{2}\mathrm{(CN)}_{3}$\cite%
{Shimizu,Kawamoto,Kurosaki}. Motivated by experiments, $U(1)$ slave-rotor
theory of Hubbard model on triangular lattice\cite{Lee} and its $SU(2)$
generalization on honeycomb lattice were formulated\cite{Hermele}. It is
predicted that quantum spin liquids may lie in the insulating side of the
metal-insulator (MI) transitions. Because the spin liquid is adjacent to the
MI transitions, people may guess it is the local charge fluctuations rather
than frustrations that disrupt spin ordering and drive the ground state to a
spin liquid. Such type of spin liquid can be described by the Hubbard model
formalism of the intermediate coupling region.

Recently, it becomes hot to use ultracold atoms as simulators of quantum
many-body systems\cite{Bloch's review}. In particular, the $\pi $-flux
Hubbard model (or the Hubbard model with $\phi $ flux) on square lattice has
been designed with ultracold atoms in an optical lattice. An artificial
magnetic field of $\pi $-flux (or $\phi $ flux) in an optical square lattice
is proposed to be realized by different approaches, such as laser assisted
tunneling method\cite{2}, laser methods by employing dark states\cite{1} or
dressing two-photon by laser fields\cite{3}. Without the nesting condition,
the \textbf{MI transition} of the $\pi $-flux Hubbard model may differ from
that of the traditional Hubbard model on square lattice. Thus, due to nodal
fermions in the non-interacting limit, it becomes an interesting issue to
learn the \textbf{MI transition} of the $\pi $-flux Hubbard model. In
addition, it is known that the insulator state of the $\pi $-flux Hubbard
model belongs to a special class of antiferromagnetic (AF) ordered state -
nodal AF insulator (NAI), an AF order with relativistic massive fermionic
excitations\cite{kou}. Another issue here is whether the nodal AF insulator
is a long range AF order or a short range one.

In the followings an O(3) non-linear $\sigma $ model ($\mathrm{NL}\sigma
\mathrm{M}$) is developed to investigate properties of NAI in the $\pi $%
-flux Hubbard model. Based on the $\mathrm{NL}\sigma \mathrm{M,}$ we will
show that a non-magnetic insulator (a short range AF order) may exist in the
NAI of 2D $\pi $-flux Hubbard model when the spin fluctuations are
considered.

\begin{figure}[tbp]
\includegraphics[width=0.43\textwidth]{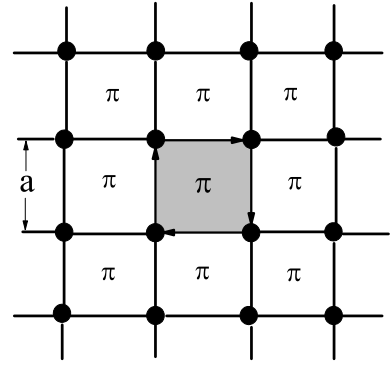}
\caption{Illustrations of a $\protect \pi $\textit{-flux} lattice. There is a
$\protect \pi $-flux phase when an atom hops aroud a plaquette (the gray
rectangle). Where $a$ is the length of the side that is chosen to be unit. }
\end{figure}

\textit{Metal-Insulator transitions of the }$\pi $\textit{-flux} \textit{%
Hubbard model} - The Hamiltonian of 2D $\pi $-flux Hubbard model is
\begin{equation}
\mathcal{H}=-\sum \limits_{\langle i,j\rangle }\left( t_{ij}\hat{c}%
_{i}^{\dagger }\hat{c}_{j}+h.c.\right) +U\sum_{i}\hat{n}_{i\uparrow }\hat{n}%
_{i\downarrow }-\mu \sum \limits_{i}\hat{c}_{i}^{\dagger }\hat{c}_{i}.
\label{mo}
\end{equation}%
Here $\hat{c}_{i}=(\hat{c}_{i\uparrow },\hat{c}_{i\downarrow })^{T}$ are
defined as electronic annihilation operators. $U$ is the on-site Coulomb
repulsion. $\mu $ is the chemical potential and at half-filling it is $\frac{%
U}{2}$. $\langle i,j\rangle $ denotes two sites on a nearest-neighbor link. $%
\hat{n}_{i\uparrow }$ and $\hat{n}_{i\downarrow }$ are the number operators
of electrons at site $i$ with up-spin and down spin, respectively. There is
a $\pi $-flux phase when an atom hops around a plaquette in a $\pi $-flux
lattice(See Fig.1). So the nearest neighbor hopping $t_{i,j}$ in a $\pi $%
-flux lattice could be chosen as\cite{hsu} $t_{i,i+\hat{x}}=t,t_{i,i+\hat{y}%
}=te^{\pm i\frac{\pi }{2}}.$

\begin{figure}[tbp]
\includegraphics[width=0.5\textwidth]{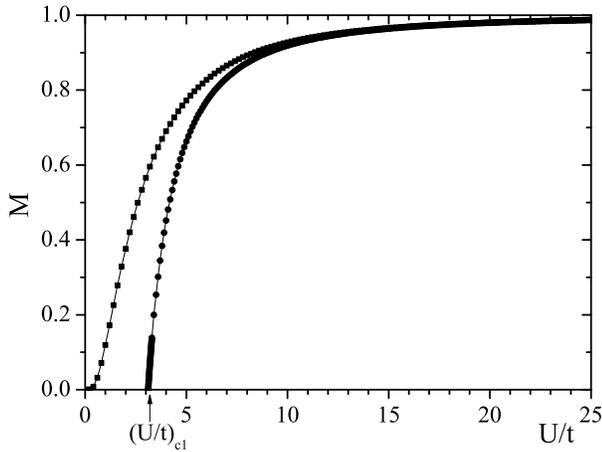}
\caption{The staggered magnetization of the $\protect \pi $\textit{-flux}
Hubbard model (circle solid line) and that of the traditional Hubbard model
(square solid line) at zero temperature. $\left( U/t\right) _{c1}\simeq 3.11$
is the critical point of the metal-insulator(MI) transition of the $\protect%
\pi $\textit{-flux} Hubbard model.}
\end{figure}

Because the Hubbard model on bipartite lattices is unstable against
antiferromagnetic (AF) instability, at half-filling, the ground state may be
an insulator with AF order (NAI). Such AF order is described by the
following mean field result
\begin{equation}
\langle \hat{c}_{i,\sigma }^{\dag }\hat{c}_{i,\sigma }\rangle =\frac{1}{2}%
\left( 1+(-1)^{i}\sigma M\right)
\end{equation}%
Here $M$ is the staggered magnetization. For the cases of spin up and spin
down, we have $\sigma =+1$ and $\sigma =-1$, respectively. Then in the mean
field theory, the Hamiltonian of the 2D $\pi $-flux Hubbard model is
obtained as%
\begin{equation}
\mathcal{H}=-\sum \limits_{\left \langle ij\right \rangle }(t_{i,j}\hat{c}%
_{i}^{\dagger }\hat{c}_{j}+h.c.)-\sum_{i}\left( -1\right) ^{i}\Delta \hat{c}%
_{i}^{\dagger }\mathbf{\sigma }_{z}\hat{c}_{i}  \label{m}
\end{equation}%
where $\Delta =\frac{UM}{2}$ leads to the energy gap of electrons and $%
\sigma _{z}$ is the Pauli matrix. After diagonalization, the spectrum of the
electrons is obtained as
\begin{equation}
E_{\mathbf{k}}=\pm \sqrt{\left \vert \xi _{\mathbf{k}}\right \vert
^{2}+\Delta ^{2}}
\end{equation}%
where $\left \vert \xi _{\mathbf{k}}\right \vert =\sqrt{4t^{2}\left( \cos
^{2}k_{x}+\cos ^{2}k_{y}\right) }$\ corresponds to the energy of free
fermions. By minimizing the free energy at temperature $T$ in the Brillouin
zone, the self-consistent equation of (\ref{m}) is reduced into
\begin{equation}
\frac{1}{N}\sum \limits_{\mathbf{k}}\frac{U}{2E_{\mathbf{k}}}\tanh (\frac{E_{%
\mathbf{k}}}{2k_{B}T})=1
\end{equation}%
where $N$ is the number of the particles.

It is well known that due to the nesting effect, there is no MI transition
of the traditional Hubbard model (arbitrary interaction will lead to a
magnetic instability). The situation is difference for the MI transition of $%
\pi $-flux Hubbard model. The MI transition of the $\pi $-flux Hubbard model
occurs at a critical value about $U/t\simeq 3.11$\cite{hsu} (See Fig.2). In
the weakly coupling limit $\left( U/t<3.11\right) $, the ground state is a
semi-metal (SM) with \emph{nodal} fermi-points\cite{kou}. In the strong
coupling region $\left( U/t>3.11\right) $, due to $M\neq 0$, the ground
state becomes an insulator with massive fermionic excitations. By contrast,
there is only the insulating phase of the traditional Hubbard model (See
Fig.2). However, the non-zero value of $M$ only means the existence of
effective spin moments. It does not necessarily imply that the ground state
of NAI is a long range AF order because the direction of the spins is chosen
to be fixed along $\mathbf{\hat{z}}$-axis in the mean field theory. Thus one
needs to examine stability of magnetic order against quantum fluctuations of
effective spin moments based on a formulation by keeping spin rotation
symmetry.

\textit{Effective nonlinear }$\sigma $\textit{\ model of spin fluctuations -
}In the following parts we will focus on the NAI state and don't consider
local charge fluctuations and the amplitude fluctuations of $M$ that are all
gapped in the region of $M\neq 0$\cite{com1}. To deal with the spin
fluctuations, we use the path-integral formulation of electrons with spin
rotation symmetry\cite{wen,dup,dupuis,Schulz,weng}. The interaction term in
Eq.(\ref{mo}) can be handled by using the SU(2) invariant
Hubbard-Stratonovich decomposition in the arbitrary on-site unit vector $%
\mathbf{\Omega }_{i}$
\begin{equation}
\hat{n}_{i\uparrow }\hat{n}_{i\downarrow }=\frac{\left( \hat{c}_{i}^{\dagger
}\hat{c}_{i}\right) ^{2}}{4}-\frac{1}{4}[\mathbf{\Omega }_{i}\mathbf{\cdot }%
\hat{c}_{i}^{\dag }\mathbf{\sigma }\hat{c}_{i}]^{2}
\end{equation}%
Here $\mathbf{\sigma =}\left( \sigma _{x},\sigma _{y},\sigma _{z}\right) $
are the Pauli matrices. By replacing the electronic operators $\hat{c}%
_{i}^{\dagger }$ and $\hat{c}_{j}$ by Grassmann variables $c_{i}^{\ast }$
and $c_{j}$, the effective Lagrangian of the 2D $\pi $-flux Hubbard model at
half filling is obtained:
\begin{equation}
\mathcal{L}_{\mathrm{eff}}=\sum_{i}c_{i}^{\ast }\partial _{\tau }c_{i}-\sum
\limits_{\left \langle ij\right \rangle }\left( t_{i,j}c_{i}^{\ast
}c_{j}+h.c.\right) -\Delta \sum_{i}c_{i}^{\ast }\mathbf{\Omega }_{i}\mathbf{%
\cdot \sigma }c_{i}.
\end{equation}%
In particular, we describe the vector $\mathbf{\Omega }_{i}$ with the
haldane's mapping:
\begin{equation}
\mathbf{\Omega }_{i}=(-1)^{i}\mathbf{n}_{i}\sqrt{1-\mathbf{L}_{i}^{2}}+%
\mathbf{L}_{i}.
\end{equation}%
Here $\mathbf{n}_{i}=(n_{i}^{x},n_{i}^{y},n_{i}^{z})$ is the N\'{e}el vector
with $\mathbf{n}_{i}^{2}=1$ and $\mathbf{L}_{i}$ is the small transverse
canting field with $\mathbf{L}_{i}\cdot $ $\mathbf{n}_{i}=0$\cite%
{dupuis,Haldane,Auerbach}.

Then we rotate $\mathbf{n}_{i}$ to $\mathbf{\hat{z}}$-axis at each site on
both sublattices by performing the following spin transformation\cite%
{wen,dup,dupuis,Schulz,weng}, $\psi _{i}=U_{i}c_{i},$ $U_{i}^{\dagger }%
\mathbf{n}_{i}\cdot \mathbf{\sigma }U_{i}=\mathbf{\sigma }_{z}$ and $%
U_{i}^{\dagger }\mathbf{L}_{i}\cdot \mathbf{\sigma }U_{i}=\mathbf{l}%
_{i}\cdot \mathbf{\sigma .}$ After the spin transformation, the effective
Hamiltonian becomes:
\begin{widetext}
\begin{equation}
\mathcal{H}_{\mathrm{eff}}=\sum \limits_{i}\psi _{i}^{\ast }a_{0}\left(
i\right) \psi _{i}-\sum \limits_{<ij>}(t_{i,j}\psi _{i}^{\ast
}e^{ia_{ij}}\psi _{j}+h.c.)-\Delta \sum \limits_{i}\psi _{i}^{\ast }[(-1)^{i}%
\mathbf{\sigma }_{z}\sqrt{1-\mathbf{l}_{i}^{2}}+\mathbf{l}_{i}\cdot \mathbf{%
\sigma ]}\psi _{i}  \label{ham}
\end{equation}%
\end{widetext}where the auxiliary gauge fields $a_{ij}=a_{ij,1}\sigma
_{x}+a_{ij,2}\sigma _{y}$ and $a_{0}\left( i\right) =a_{0,1}\left( i\right)
\sigma _{x}+a_{0,2}\left( i\right) \sigma _{y}\ $are defined by $%
e^{ia_{ij}}=U_{i}^{\dagger }U_{j}$ and $a_{0}\left( i\right) =U_{i}^{\dagger
}\partial _{\tau }U_{i}.$ In terms of the mean field result, $M=\left(
-1\right) ^{i}\langle \psi _{i}^{\ast }\mathbf{\sigma }_{z}\psi _{i}\rangle ,
$ we obtain the effective Hamiltonian:
\begin{widetext}
\begin{eqnarray}
\mathcal{H}_{\mathrm{eff}} &\simeq &\sum \limits_{i}\psi _{i}^{\ast
}[a_{0}\left( i\right) -\Delta \mathbf{\sigma \cdot
\mathbf{l}_{i}}]\psi _{i}-\Delta \sum \limits_{i}(-1)^{i}\psi
_{i}^{\ast }\sigma _{z}\psi _{i} -\sum \limits_{\left \langle
ij\right \rangle }[t_{i,j}\psi _{i}^{\ast
}(1+ia_{ij})\psi _{j}+h.c.]+\Delta M\sum \limits_{i}\frac{\mathbf{l}_{i}^{2}%
}{2}
\end{eqnarray}%
\end{widetext}In this equation we have used the approximations $\sqrt{1-%
\mathbf{l}_{i}^{2}}\simeq 1-\frac{\mathbf{l}_{i}^{2}}{2}$ and $%
e^{ia_{ij}}\simeq 1+ia_{ij}$.

In the next step, we integrate the gapped fermion fields and get the quadric
terms of $[a_{0}\left( i\right) -\Delta \mathbf{\sigma \cdot l}_{i}]$ and $%
a_{ij}$. Then the effective action becomes:%
\begin{equation}
\mathcal{S}_{\mathrm{eff}}=\frac{1}{2}\int_{0}^{\beta }d\tau
\sum_{i}[-4\varsigma (a_{0}\left( i\right) -\Delta \mathbf{\sigma \cdot l}%
_{i})^{2}+4\rho _{s}a_{ij}^{2}+\frac{2\Delta ^{2}}{U}\mathbf{l}_{i}^{2}]
\label{efff}
\end{equation}%
where the parameters $\rho _{s}$ and $\varsigma $ are derived from the
following two equations\cite{det}:
\begin{eqnarray}
\rho _{s} &=&\frac{1}{N}\sum \limits_{\mathbf{k}}\frac{\epsilon ^{2}}{%
2(\left \vert \xi _{\mathbf{k}}\right \vert ^{2}+\Delta ^{2})^{\frac{3}{2}}},
\\
\varsigma  &=&\frac{1}{N}\sum \limits_{\mathbf{k}}\frac{\Delta ^{2}}{%
4(\left \vert \xi _{\mathbf{k}}\right \vert ^{2}+\Delta ^{2})^{\frac{3}{2}}}.
\end{eqnarray}%
and the corresponding coefficient $\epsilon ^{2}$ is given as:%
\begin{widetext}
\begin{eqnarray}
\epsilon ^{2}=t^{2}[\cos \left( 2k_{x}\right) \left( \Delta
^{2}+8t^{2}+4t^{2}\cos \left( 2k_{y}\right) \right) +\Delta
^{2}+3t^{2}+t^{2}\cos \left( 4k_{x}\right) ]
\end{eqnarray}%
\end{widetext}

To learn the properties of the low energy physics, we study the continuum
theory of the effective action in Eq.(\ref{efff}). In the continuum limit,
we denote $\mathbf{n}_{i}$, $\mathbf{l}_{i}$, $ia_{ij}\simeq U_{i}^{\dagger
}U_{j}-1$ and $a_{0}\left( i\right) =U_{i}^{\dagger }\partial _{\tau }U_{i}$
by $\mathbf{n}(x,y)$, $\mathbf{l}(x,y)$, $U^{\dagger }\partial _{x}U$ (or $%
U^{\dagger }\partial _{y}U$) and $U^{\dagger }\partial _{\tau }U,$
respectively. From the relations between $U^{\dagger }\partial _{\mu }U$ and
$\partial _{\mu }\mathbf{n,}$
\begin{eqnarray}
a_{\tau }^{2} &=&a_{\tau ,1}^{2}+a_{\tau ,2}^{2}=-\frac{1}{4}(\partial
_{\tau }\mathbf{n})^{2},\text{ }\tau =0, \\
a_{\mu }^{2} &=&a_{\mu ,1}^{2}+a_{\mu ,2}^{2}=\frac{1}{4}(\partial _{\mu }%
\mathbf{n})^{2},\text{ }\mu =x,y, \\
\mathbf{a}_{0}\mathbf{\cdot l} &\mathbf{=}&\mathbf{-}\frac{i}{2}\left(
\mathbf{n}\times \partial _{\tau }\mathbf{n}\right) \cdot \mathbf{l,}
\end{eqnarray}%
the continuum formulation of the action in Eq.(\ref{efff}) turns into
\begin{widetext}
\begin{equation}
\mathcal{S}_{\mathrm{eff}}=\frac{1}{2}\int_{0}^{\beta }d\tau \int d^{2}%
\mathbf{r}[\varsigma (\partial _{\tau }\mathbf{n)}^{2}+\rho _{s}\left(
\mathbf{\bigtriangledown n}\right) ^{2}-4i\Delta \varsigma \left( \mathbf{n}%
\times \partial _{\tau }\mathbf{n}\right) \cdot \mathbf{l}+(\frac{2\Delta
^{2}}{U}-4\Delta ^{2}\varsigma )\mathbf{l}^{2}]
\end{equation}%
\end{widetext}where the vector $\mathbf{a}_{0}$ is defined as $\left(
a_{0,1},a_{0,2},0\right) .$

Finally we integrate the transverse canting field $\mathbf{l}$ and obtain
the effective $\mathrm{NL}\sigma \mathrm{M}$ of the $\pi $-flux Hubbard
model as
\begin{equation}
\mathcal{S}_{\mathrm{eff}}=\frac{1}{2g}\int_{0}^{\beta }d\tau \int d^{2}r[%
\frac{1}{c}\left( \partial _{\tau }\mathbf{n}\right) ^{2}+c\left( \mathbf{%
\bigtriangledown n}\right) ^{2}]\text{ }  \label{non}
\end{equation}%
with a constraint $\mathbf{n}^{2}=1.$ The coupling constant $g$ and spin
wave velocity $c$ are defined as
\begin{figure}[tbp]
\includegraphics[width=0.5\textwidth]{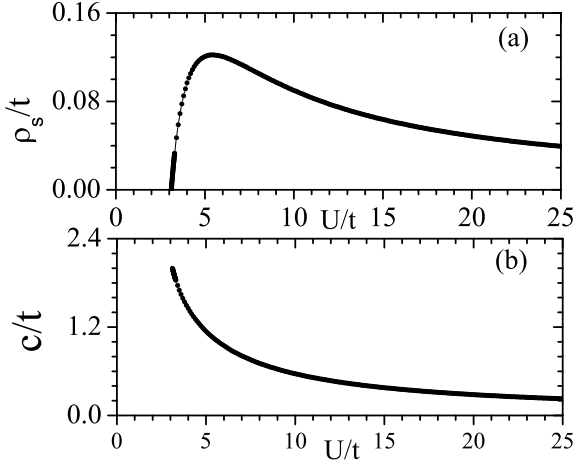}
\caption{The spin stiffness $\protect \rho _{s}$ and the spin wave velocity $c
$ of the $\protect \pi $-flux Hubbard model.}
\end{figure}
\begin{eqnarray}
g &=&\frac{c}{\rho _{s}} \\
c^{2} &=&\frac{\rho _{s}}{\chi ^{\perp }}
\end{eqnarray}%
Here $\rho _{s}$ is the spin stiffness and $\chi ^{\perp }=(\frac{1}{%
\varsigma }-2U)^{-1}$ is the transverse spin susceptibility.

The numerical results of $\rho _{s}$ and $c$ of the $\pi $-flux Hubbard
model are illustrated in Fig.3, where one can find that $\rho
_{s}=0.03936t=0.2460J$, $c=0.226278t=1.41424J$ in the strong coupling limit$%
\left( U=25t\right) $ match the earlier results $\rho _{s}=\frac{J}{4},$ $c=%
\sqrt{2}J$ ($J=\frac{4t^{2}}{U}$) that are obtained from the Heisenberg
model \cite{Schrieffer,Singh,Chubukov,man}.

In addition, we need to determine another important parameter - the cutoff $%
\Lambda $. On the one hand, the effective $\mathrm{NL}\sigma \mathrm{M}$ is
valid within the energy scale of Mott gap, $2\Delta =UM.$ On the other hand,
the lattice constant is a natural cutoff. Thus the cutoff is defined as the
following equation $\Lambda =\min (1,\frac{2\Delta }{c})$\cite{dupuis}.

\textit{Magnetic properties of the nodal AF insulator : }In this section we
will use the effective $\mathrm{NL}\sigma \mathrm{M}$ to study the magnetic
properties of the insulator state. The Lagrangian\ of $\mathrm{NL}\sigma
\mathrm{M}$ with a constraint ($\mathbf{n}^{2}=1$) by a Lagrange multiplier $%
\lambda $ becomes
\begin{equation}
\mathcal{L}_{\mathrm{eff}}=\frac{1}{2cg}\left[ \left( \partial _{\tau }%
\mathbf{n}\right) ^{2}+c^{2}\left( \mathbf{\bigtriangledown n}\right)
^{2}+i\lambda (1-\mathbf{n}^{2})\right]  \label{cons}
\end{equation}%
where $i\lambda =m^{2}$ and $m$ is the mass gap of the spin fluctuations.

At finite temperature, by rescaling the field $n\rightarrow \sqrt{N}n$\ and
using the large-N approximation, the solution of $n_{0}=\left \langle
\mathbf{n}\right \rangle $\ is always zero that is consistent to the
Mermin-Wigner theorem. From Eq.(\ref{cons}), we may get the solution of $m$
as
\begin{equation}
m=2T\sinh ^{-1}[e^{-\frac{2\pi c}{gT}}\sinh (\frac{c\Lambda }{2k_{\mathrm{B}%
}T})].
\end{equation}%
\begin{figure}[tbp]
\includegraphics[width=0.5\textwidth]{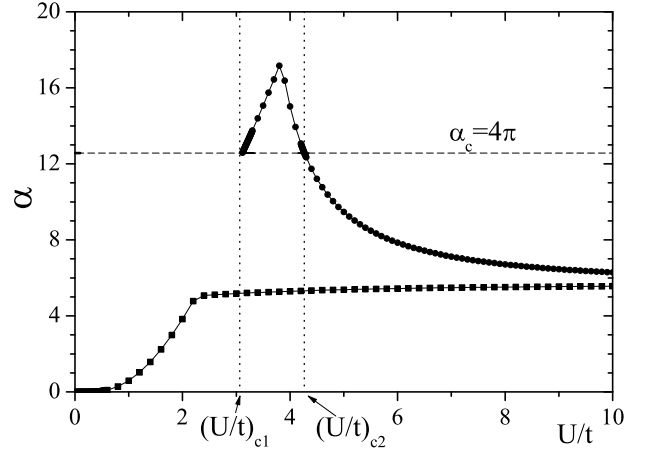}
\caption{The dimensionless coupling constant $\protect \alpha $ of the $%
\protect \pi $-flux Hubbard model (cicle solid line) and that of the
traditional Hubbard model (square solid line). There are three regimes,
semimetal(SM), quantum disordered(QD), antiferromagnetic(AF), separated by
two critical points $\left( U/t\right) _{c1}\simeq 3.11$, $\left( U/t\right)
_{c2}\simeq 4.26,$ respectively. There is only the AF regime on the
traditional square lattice.}
\end{figure}
\  \  \

At zero temperature, the solutions of $n_{0}$ and $m$ of Eq.(\ref{cons}) are
determined by the dimensionless coupling constant $\alpha =g\Lambda .$\ In
particular, there exists a critical point $\alpha _{c}=4\pi $ $($ or $g_{c}=%
\frac{4\pi }{\Lambda })$ : For the case of $\alpha <4\pi ,$ we get solutions
of $n_{0}$ and $m$ as $n_{0}=(1-\frac{g}{g_{c}})^{1/2}$ and $m=0$. For the
case of $\alpha >4\pi ,$ we get solutions of $n_{0}$ and $m$ as $n_{0}=0$
and $m=4\pi c(\frac{1}{g_{c}}-\frac{1}{g})$. So we calculate the
dimensionless coupling constant $\alpha =g\Lambda $ of the $\pi $-flux
Hubbard model and show results in Fig.4. The quantum critical point
corresponding to $\alpha _{c}=4\pi $ turns into $U/t\simeq 4.26$ which
devides the NAI state into two phases - a quantum disordered state (QD) in
the region of $3.11<U/t<4.26$ and a long range AF order in the region of $%
U/t>4.26$. The results show sharp contrast to those from the traditional
Hubbard model, where the dimensionless coupling constant is always smaller
than $\alpha _{c}=4\pi $.

In the region of $U/t>4.26$ (where $\alpha <\alpha _{c}$), at low
temperature the mass gap $m$ of spin fluctuations is determined by
\begin{equation}
m\simeq 2k_{\mathrm{B}}T\exp [-\frac{2\pi c}{k_{\mathrm{B}}T}(\frac{1}{g}-%
\frac{1}{g_{c}})].
\end{equation}%
Because the energy scale of the mass gap $m$ is always much smaller than the
temperature, \textit{i.e.}, $m\ll k_{\mathrm{B}}T$ (or $\omega _{n}$),
quantum fluctuations become negligible in a sufficiently long wavelength and
low energy regime $\left( m<\left \vert c\mathbf{q}\right \vert <k_{\mathrm{B%
}}T\right) .$ Thus in this region one may only consider the purely static
(semiclassical) fluctuations. The effective Lagrangian of the NL$\sigma $M
then becomes%
\begin{equation}
\mathcal{L}=\frac{\tilde{\rho}_{s}}{2}\left( \mathbf{\bigtriangledown n}%
\right) ^{2}
\end{equation}%
where $\tilde{\rho}_{s}=c\left( \frac{1}{g}-\frac{1}{g_{c}}\right) $ is the
renomalized spin stiffness. At zero temperature, the mass gap vanishes (See
Fig.5(a)) which means that long range AF order appears. To describe the long
range AF order, we introduce a spin order parameter\cite{cha,chu,sech}
\begin{equation}
\mathcal{M}_{0}=\frac{M}{2}n_{0}=\frac{M}{2}(1-\frac{g}{g_{c}})^{1/2}
\end{equation}%
As shown in Fig.5(b), the ground state of long range AF ordered phase has a
finite spin order parameter. In addition, in the strong coupling limit, $%
U/t\rightarrow \infty $, the values naturally match the results derived from
the Heisenberg model mapped from the $\pi $-flux Hubbard model.

\begin{figure}[tbp]
\includegraphics[width=0.5\textwidth]{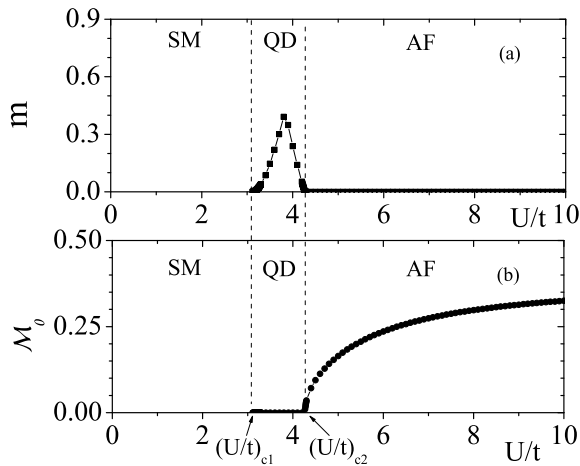}
\caption{The mass gap $m$ of the spin fluctuations and the ordered spin
moment $\mathcal{M}_{0}$ of the $\protect \pi $-flux Hubbard model at zero
temperature. There are three regimes, semimetal(SM), quantum disordered(QD),
antiferromagnetic(AF), separated by two critical points $\left( U/t\right)
_{c1}\simeq 3.11$, $\left( U/t\right) _{c2}\simeq 4.26,$ respectively. }
\end{figure}

In the region of $3.11<U/t<4.26$ $($where $\alpha >\alpha _{c}),$ there is a
finite mass gap of spin fluctuations, $m=4\pi c(\frac{1}{g_{c}}-\frac{1}{g})$
at zero temperature (See Fig.5(a)). Therefore, the ground state of the
insulator in this region is not a long range AF order. Instead, it is a
quantum disordered state (or non-magnetic insulator state) with zero spin
order parameter $\mathcal{M}_{0}=0$ (See Fig.5(b)). The existence of a
non-magnetic insulator state provides an alternative candidate for finding
\emph{spin liquid} state.

Obviously, such type of quantum non-magnetic insulator on bipartite lattices
is induced neither by geometry frustrations regarded as the examples in
varied spin models nor by the local charge fluctuations with finite energy
gap. What is the physics origin of this quantum non-magnetic state? The key
point is that, \emph{due to the special electron dispersion (the existence
of nodal fermions for non-interacting case) the coupling constant }$g$\emph{%
\ is almost proportional to }$\frac{1}{M}$\emph{\ near the MI transition}
(See Fig.6). Hence the non-magnetic state originates from quantum spin
fluctuations of relatively small effective spin-moments, $M\rightarrow 0$.
\begin{figure}[tbp]
\includegraphics[width=0.5\textwidth]{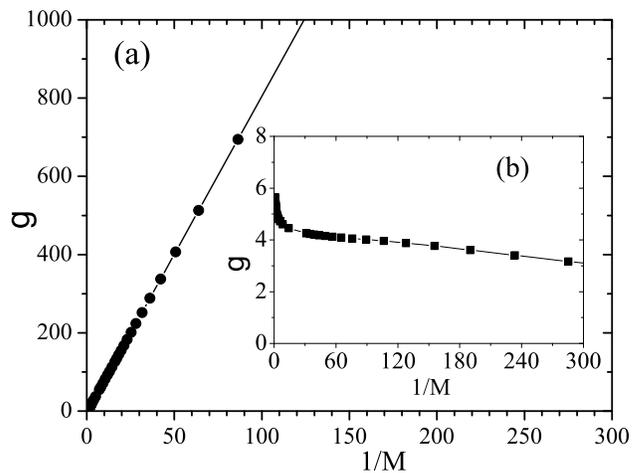}
\caption{Illustrations of the relations between the coupling constant $g$\
and the staggered magnetization $M$ of the $\protect \pi $-flux Hubbard
model(cicle solid line) and the traditional Hubbard model (square solid line
in inset (b)).}
\end{figure}

Let us compare the properties of the insulator state in $\pi $-flux Hubbard
model and those in the traditional Hubbard model. For the traditional
Hubbard model on square lattice, due to the nesting effect, there is no MI
transition at finite $U$ and the insulator state here doesn't belong to NAI.
In the $U/t\rightarrow 0$\ limit, the coupling constant $g$ is not
proportional to $\frac{1}{M}$(See Fig.6). Instead, $g$ is about $g\sim \frac{%
2}{\sqrt{\pi }}(\frac{U}{t})^{1/4}$ that becomes smaller and turns into zero
the weakly coupling limit(See more details in Ref.\cite{dupuis}). So the
quantum fluctuations of the effective spin-moments are suppressed. Using the
\textrm{NL}$\sigma $\textrm{M} formulation, due to $g<g_{c}$ (See Fig.4),
the ground state of the Hubbard model on square lattice always has a long
range AF\ order.

\textit{Conclusion - }In this paper, to deal with the spin fluctuations, we
use the path-integral formulation of electrons with spin rotation symmetry
and then the effective $\mathrm{NL}\sigma \mathrm{M}$ is obtained to
describe the NAI state of the $\pi $-flux Hubbard model. We calculate the
spin stiffness, the transverse spin susceptibility, the spin wave velocity
and the coupling constant $g$. In the strong coupling limit ($U/t\rightarrow
\infty $), our results of spin velocity and spin order parameter agree with
the results obtained from earlier calculations of the traditional Hubbard
model. However, we find that the coupling constant $g$ in the NAI state of
the $\pi $-flux Hubbard model shows different behaviors to that in the
insulator state of the traditional Hubbard model. In particular, a quantum
non-magnetic insulator state $\left( 3.11<U/t<4.26\right) $ is explored near
the MI transition that corresponds to the strong coupling region of the
effective $\mathrm{NL}\sigma \mathrm{M,}$ $g>g_{c}$. Such type of quantum
non-magnetic insulator in bipartite lattices is driven by quantum spin
fluctuations of relatively small effective spin-moments.

Such non-magnetic insulator state is different \textbf{from that proposed}
in organic material $\mathrm{\kappa -(BEDT-TTF)}_{2}\mathrm{CU}_{2}\mathrm{%
(CN)}_{3}$ by $U(1)$ slave-rotor theory in Ref.\cite{Lee}. Firstly, the
non-magnetic insulator state is really a short range AF insulator (although
we don't know its exact nature) followed by a long range AF order with
increasing $U;$ however, the spin liquid state in Ref.\cite{Lee} is really a
$U(1)$ spin liquid with spinon fermi surface, of which no long range AF
order appears with increasing $U$. Secondly, the local charge fluctuations
play important role in the slave-rotor theory; in contrast, the local charge
fluctuations are irrelevant here. Therefore, our results illustrate a new
candidate for finding spin liquid state.

An interesting issue is the nature of the non-magnetic insulator. Is it a
valence-band crystal \cite{read}, or algebra spin liquid state\cite{wen,wen1}%
, ...? In addition, another issue is whether there exists non-magnetic
insulator state of the Hubbard model in honeycomb lattice, of which there
also exist nodal fermions. These issues are beyond the scope of the present
work and will be left for a future study.

The authors acknowledge stimulating discussions with Z.Y. Weng, T. Li, N.H.
Tong, F. Yang, P. Ye. This research is supported by NCET, NFSC Grant no.
10574014, 10874017.


\begin{thebibliography}{99}
\bibitem{com1} When the mean field value of $M$ almost vanishes, approaching
the critical point ($U/t\simeq 3.11$), due to the strong charge fluctuations
and the strong amplitude fluctuations of $M$, $\mathrm{NL}\sigma \mathrm{M}$
cannot be used and our results cannot be reliable.
\end{thebibliography}
\end{document}